# Magnon magic angles and tunable Hall conductivity in 2D twisted ferromagnetic bilayers

Doried Ghader

College of Engineering and Technology, American University of the Middle East, Kuwait

**Abstract.** Twistronics is currently one of the most active research fields in condensed matter physics, following the discovery of correlated insulating and superconducting phases in twisted bilayer graphene (tBLG). Here, we present a magnonic analogue of tBLG. We study magnons in twisted ferromagnetic bilayers (tFBL), including exchange and Dzyaloshinskii-Moriya interactions (DMI). For negligible DMI, tFBL presents discrete magnon magic angles and flat moiré minibands analogous to tBLG. The DMI, however, changes the picture and renders the system much more exotic. The DMI in tFBL induces a rich topological magnon band structure for any twist angle. The twist angle turns to a control knob for the magnon Hall and Nernst conductivities. Gapped flat bands appear in a continuum of magic angles in tFBL with DMI. In the lower limit of the continuum, the band structure reconstructs to form bundles of topological flat bands. The luxury of twist-angle control over band gaps, topological properties, number of flat bands, Hall and Nernst conductivities renders tFBL a novel device from fundamental and applied perspectives.

**Introduction.** Two-dimensional (2D) materials with intrinsic magnetism has recently been realized [1, 3], opening new horizons in 2D material research [4 – 26]. In these bosonic Dirac materials, magnetic anisotropy is found to overcome thermal fluctuations and stabilize the magnetic order at finite temperatures. The exotic physics in 2D magnetic systems attracted important attention in search for novel nanomagnetic quantum devices.

To a large extent, the theoretical investigation and experimental realization of bosonic Dirac materials was motivated by their fermionic counterparts. Research on graphene demonstrates that the electronic properties in bilayers change drastically compared to single layer graphene [27 - 29]. A particularly interesting class of bilayer graphene is the twisted bilayer graphene (tBLG), presenting moiré Bloch bands as a result of the twist. tBLG was found to present fascinating electronic and optical properties, giving rise to novel physics that is completely absent in AB stacked bilayer graphene [30-39]. The twist angle reconstructs the electronic structure, realizing flat moiré superlattice minibands at discrete magic angles. Superconductivity was observed at magic angles in tBLG [38, 39] which triggered unprecedented interest in 2D moiré materials [40-47]. Numerous fermionic 2D heterostructure are currently under intensive investigation for superconducting, correlation and topological features.

Magnons in 2D magnetic materials mimic electrons in 2D fermionic systems [15]. For example, the exchange magnon spectrum in a 2D honeycomb ferromagnet is qualitatively identical to the electronic structure in graphene. Moreover, 2D and quasi-2D quantum magnets with Dzyaloshinskii-Moriya (DM) spin-orbit interaction can host topological magnon bands [6 - 8, 12, 14, 18, 21], similar to their fermionic counterparts. The topological nature of the magnon spectrum in 2D magnets can be confirmed via the thermal magnon Hall response. The magnon Hall conductivity in honeycomb ferromagnets with DMI was investigated in monolayers [7] and AB stacked bilayers [8].

Given the remarkable analogy between graphene and honeycomb ferromagnets, it is reasonable to propose tFBL with ferromagnetic interlayer coupling (e.g. $CrBr_3$ and $Cr_2Ge_2Te_6$) as potential magnonic analogues for tBLG. Bilayers formed of 2D magnetic materials are van der Waals materials with a weak interlayer exchange coupling [8, 10, 17, 18]. The ferromagnetic interlayer exchange in the proposed tFBL thus mimics the weak interlayer hopping in tBLG. The arguments in the Bistritzer - MacDonald approach for tBLG [32] can hence be implemented to develop the tFBL spin wave theory. In the absence of the DMI, the magnons transport properties in tFBL are found to mimic their electronic counterparts. The DMI enriches the topology in the system and induces exciting new physics. Unlike tBLG, its magnetic twin with DMI presents a continuum of magic angles and topological flat bands bundle. The magnon bands Berry curvatures and Chern numbers are sensitive to the twist angle and the DMI strength. The thermal magnon Hall and Nernst conductivities induced by the multiple topological flat bands show a complex and exotic response to the twist angle. The twist angle can hence be used as a control knob for these topological responses, which is not possible in tBLG.

**Model Hamiltonian for tFBL.** We start with a ferromagnetic honeycomb monolayer as in Fig.1a. For an A-site, the nearest and next nearest vectors are denoted $\vec{\delta}_i^A$ and $\vec{\gamma}_j$. Vectors $\vec{\gamma}_j$ also serve the B-sublattice, whereas $\vec{\delta}_i^B = -\vec{\delta}_i^A$. We also define the lattice constant $a$ as the $A - A$ (or $B - B$) distance whereas the nearest neighbor distance is $d = a/\sqrt{3}$.

In Supplementary Notes 1, we revise the spin wave theory in ferromagnetic honeycomb monolayers, including nearest neighbor exchange and next nearest neighbor DM interactions. We derive the monolayer (ML) Hamiltonian, $\mathcal{H}_{ML}$, as

$$\mathcal{H}_{ML}(\vec{k}) = JM \begin{pmatrix} 3 - \frac{iD}{J}f_D(\vec{k}) & -f(\vec{k}) \\ -f^*(\vec{k}) & 3 + \frac{iD}{J}f_D(\vec{k}) \end{pmatrix}$$

with

$$f(\vec{k}) = e^{ik_y\frac{a}{\sqrt{3}}} + 2e^{-i\frac{\sqrt{3}a}{6}k_y}\cos\left(\frac{a}{2}k_x\right)$$

$$f_D(\vec{k}) = 4\,i\,sin\left(\frac{a}{2}k_x\right)\cos\left(\frac{\sqrt{3}a}{2}k_y\right) - 2\,i\,sin(k_x a)$$

The parameters $J$ and $D$ denote the in-plane exchange and DMI coefficients respectively. $M$ is the $z-$component of the magnetization.

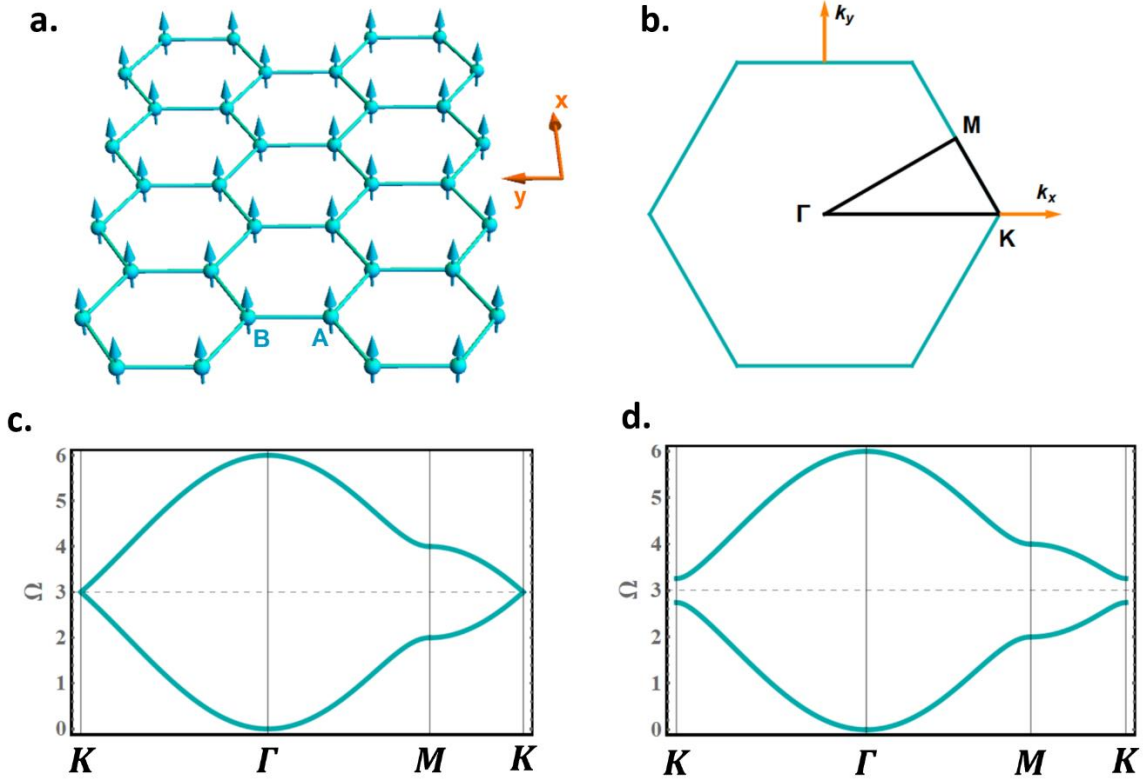

**Figure 1:** (a) Schematic representation of a single ferromagnetic honeycomb sheet. (b) The corresponding Brillouin zone and high symmetry axes. (c) and (d) show the magnon dispersion curves along the high symmetry axes for $D = 0$ and $D = 0.05J$ respectively.

Similar to graphene tight-binding Hamiltonian, $\mathcal{H}_{ML}(\vec{k})$ can be expanded near $K$ and $K' = -K$ valleys in the form of Dirac Hamiltonians,

$$\mathcal{H}_{ML}^{K}(\vec{K} + \vec{q}) = 3JMI_2 + 3\sqrt{3}DM\sigma_z + v|\vec{q}|\begin{pmatrix} 0 & e^{-i\theta_{\vec{q}}} \\ e^{i\theta_{\vec{q}}} & 0 \end{pmatrix}$$

$$\mathcal{H}_{ML}^{-K}(-\vec{K} + \vec{q}) = 3JMI_2 - 3\sqrt{3}DM\sigma_z - v|\vec{q}|\begin{pmatrix} 0 & e^{i\theta_{\vec{q}}} \\ e^{-i\theta_{\vec{q}}} & 0 \end{pmatrix}$$

with $v = \frac{1}{2}\sqrt{3}JMa$, $\vec{K} = \frac{4\pi}{3a}\hat{y}$, and $\vec{\sigma} = \sigma_x\hat{x} + \sigma_y\hat{y}$. The matrices $\sigma_i$ are the Pauli matrices while $I_2$ is the $2 \times 2$ identity matrix. $\theta_{\vec{q}}$ is the angle between momentum $\vec{q}$ and the $x-$axis in the momentum space.

Fig.1c and Fig.1d present the magnon spectra, along the high symmetry axes (Fig.1b), for $D = 0$ and $D = 0.05J$ respectively. For negligible DMI, the magnons act as massless Dirac quasi-particles near $K$, with linear dispersions. Similar to tBLG, magnon flat bands are expected in tFBL, as a result of the band repulsion effect between the overlapping Dirac cones from different layers. In the presence of the DMI, the Dirac magnons acquire mass and the magnon spectrum is gapped throughout the BZ. The Dirac cones are absent in this case and the band repulsion effect is expected to induce new dispersion profiles that are absent in tBLG.

Consider next a ferromagnetic bilayer in the AB configuration. Sites in layers $l = 1, 2$ are denoted $A_l$ and $B_l$. In the AB stacking, the constant ferromagnetic interlayer exchange coefficient, $J_\perp$, is considered between $A_1 - B_2$ dimers and neglected elsewhere. To form the tFBL, we translate layer 2 by a vector $\vec{\tau}_0 = (\tau_{0x}, \tau_{0y})$ and then rotate layers 1 and 2 in opposite directions. To be specific, layer 1 and 2 are rotated by $\theta/2$ in clockwise and anticlockwise directions respectively. The distance dependent interlayer coefficient is assumed ferromagnetic and the DMI is assumed weak. The system is in a collinear ferromagnetic ground state as in Fig.2.

We write a semi-classical Heisenberg Hamiltonian $\mathcal{H}_T$ for the tFBL, including nearest neighbor exchange and next nearest neighbors DMI as follows

$$\mathcal{H}_T = -J\sum_{l,\vec{\delta}_i^A} \vec{S}^{A_l}(\vec{R}_{A_l}, t).\vec{S}^{B_l}(\vec{R}_{A_l} + \vec{\delta}_i^A, t) - \sum_{\alpha,\beta} J_\perp(\vec{R}_{\alpha_1}, \vec{R}_{\beta_2})\, \vec{S}^{\alpha_1}(\vec{R}_{\alpha_1}, t).\vec{S}^{\beta_2}(\vec{R}_{\beta_2}, t)$$

$$+ \sum_{\alpha,l,\vec{\gamma}_j} \vec{D}(\vec{R}_{\alpha_l}, \vec{R}_{\alpha_l} + \vec{\gamma}_j).\left[\, \vec{S}^{\alpha_l}(\vec{R}_{\alpha_l}, t) \times \vec{S}^{\alpha_l}(\vec{R}_{\alpha_l} + \vec{\gamma}_j, t)\right]$$

(1)

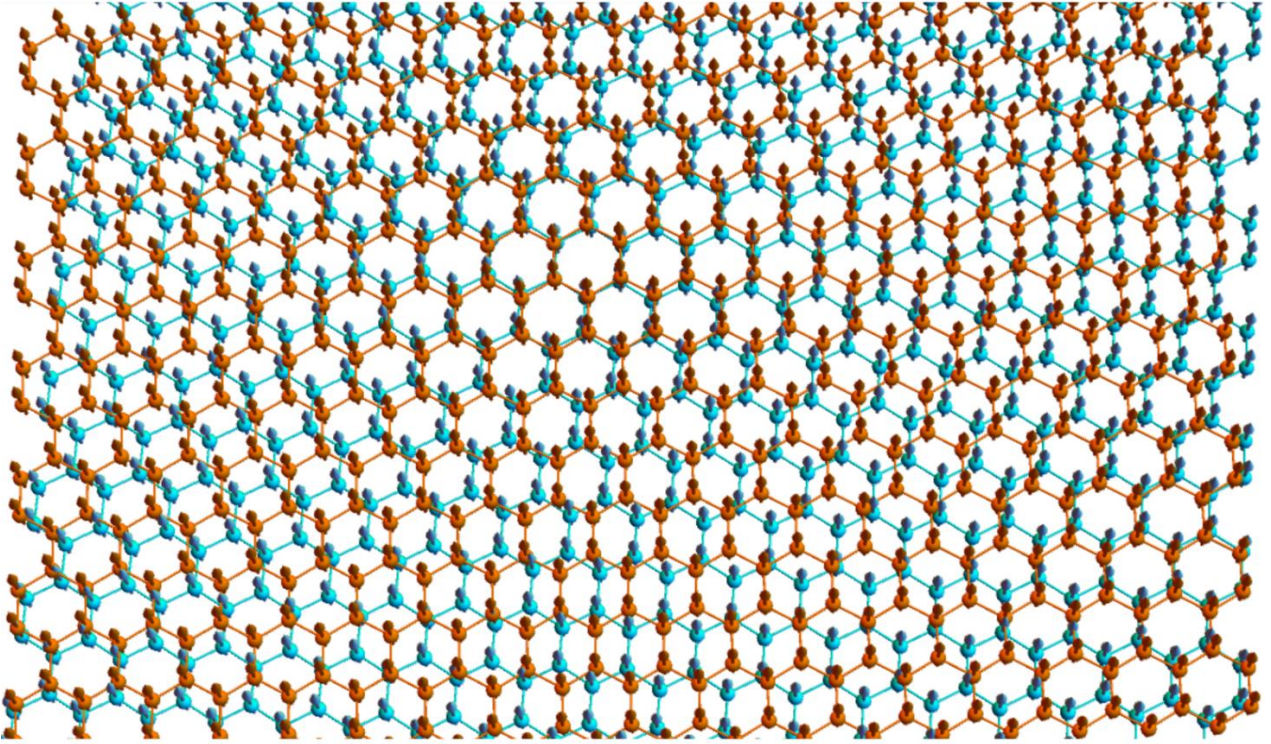

**Figure 2:** Schematic representation of tFBL for $\theta = 5°$ and $\vec{\tau}_0 = (0, d)$.

Index $l$ is summed over 1 and 2 while each of $\alpha$ and $\beta$ runs over $A$ and $B$ sites. $\vec{S}(\vec{R}, t)$ denotes the spin on site $\vec{R}$ at time $t$. $J_\perp(\vec{R}_{\alpha_1}, \vec{R}_{\beta_2})$ is the distance dependent interlayer exchange coefficient between spins at sites $\alpha_1$ and $\beta_2$. $\vec{D}$ is the alternating next nearest-neighbor DMI vector, with $\vec{D}(\vec{r}, \vec{r} + \vec{\gamma}_j) = \pm D\hat{z}$. The parameter $D$ denotes the strength of the DMI. The first, second and third terms in $\mathcal{H}_T$ hence account for the intralayer exchange, interlayer exchange and DM interactions respectively. A less compact expression for $\mathcal{H}_T$ is presented in Supplementary Note 2.

**Spin dynamics in tFBL.** The DMI term in $\mathcal{H}_T$ can be rewritten in terms of a scalar product [26] which unifies the treatment of the exchange and the DMI parts of $\mathcal{H}_T$ (Supplementary Notes 2). The effective exchange fields $\vec{H}^{\alpha_l}$ acting on the sublattice magnetizations $\vec{M}^{\alpha_l}$ can then be derived from the Heisenberg Hamiltonian as [23-26, 48-54]

$$\vec{H}^{\alpha_l}(\vec{R}_{\alpha_l}, t) = -J \sum_{\vec{\delta}_i^\alpha} \vec{M}^{\bar{\alpha}_l}(\vec{R}_{\alpha_l} + \vec{\delta}_i^\alpha, t) + \sum_{\vec{\gamma}_j} D_z(\vec{R}_{\alpha_l}, \vec{R}_{\alpha_l} + \vec{\gamma}_j) \vec{M}_D^{\alpha_l}(\vec{R}_{\alpha_l} + \vec{\gamma}_j, t)$$

$$- \sum_{\vec{R}_{\alpha_{\bar{l}}}} J_\perp(\vec{R}_{\alpha_l}, \vec{R}_{\alpha_{\bar{l}}}) \, \vec{M}^{\alpha_{\bar{l}}}(\vec{R}_{\alpha_{\bar{l}}}, t) - \sum_{\vec{R}_{\bar{\alpha}_{\bar{l}}}} J_\perp(\vec{R}_{\alpha_l}, \vec{R}_{\bar{\alpha}_{\bar{l}}}) \, \vec{M}^{\bar{\alpha}_{\bar{l}}}(\vec{R}_{\bar{\alpha}_{\bar{l}}}, t)$$

$$(2)$$

where we have used the convention that if $\alpha = A$ then $\bar{\alpha} = B$ and vice versa. Same convention assumed for $l$ and $\bar{l}$. We also introduce the vector $\vec{M}_D^{\alpha_l} = M_{y^l}^{\alpha_l} \hat{x} - M_x^{\alpha_l} \hat{y}$ to simplify the expression of $\vec{H}^{\alpha_l}$.

The spin dynamics in tFBL are governed by the Landau-Lifshitz (LL) equations of motion, $\partial_t \vec{M}^{\alpha_l} = \vec{M}^{\alpha_l} \times \vec{H}^{\alpha_l}$. Detailed development of these equations is presented in Supplementary Notes 3. Interestingly, the interlayer coefficients in the LL equations are found to be qualitatively identical to those encountered in the electronic theory for tBLG. These are hence treated using the Bistritzer - MacDonald continuum approach [32], valid for commensurate and incommensurate structures at small twist angles ($\theta \leq 10^{\circ}$). The spin wave theory, however, yields intralayer terms that are absent in the electronic theory of tBLG. Nevertheless, the main ideas of the Bistritzer - MacDonald approach can still be applied to evaluate these terms. Details are presented in Supplementary Notes 3. In conclusion, the $K$ −valley LL equations (near $K_l$ and $K_{\bar{l}}$) reduce to

$$\Omega\, u_{A_1}(\vec{K}_1 + \vec{q}) = \left[\Omega_0 + 3\sqrt{3}\frac{D}{J}\right] u_{A_1}(\vec{K}_1 + \vec{q}) + \frac{\sqrt{3}a}{2}|\vec{q}|e^{-i(\theta_q - \theta/2)}u_{B_1}(\vec{K}_1 + \vec{q})$$

$$- \frac{J_\perp}{3J}\left[u_{A_2}(\vec{K}_2 + \vec{q} + \vec{q}_b) + e^{i\varphi}u_{A_2}(\vec{K}_2 + \vec{q} + \vec{q}_{Jr}) + e^{-i\varphi}u_{A_2}(\vec{K}_2 + \vec{q} + \vec{q}_{Jl})\right]$$

$$- \frac{J_\perp}{3J}\left[u_{B_2}(\vec{K}_2 + \vec{q} + \vec{q}_b) + u_{B_2}(\vec{K}_2 + \vec{q} + \vec{q}_{Jr}) + u_{B_2}(\vec{K}_2 + \vec{q} + \vec{q}_{Jl})\right]$$

$$(3a)$$

$$\Omega\, u_{B_1}(\vec{K}_1 + \vec{q}) = \left[\Omega_0 - 3\sqrt{3}\frac{D}{J}\right] u_{B_1}(\vec{K}_1 + \vec{q}) + \frac{\sqrt{3}a}{2}|\vec{q}|e^{i(\theta_q - \theta/2)}u_{A_1}(\vec{K}_1 + \vec{q})$$

$$- \frac{J_\perp}{3J}\left[u_{A_2}(\vec{K}_2 + \vec{q} + \vec{q}_b) + e^{-i\varphi}u_{A_2}(\vec{K}_2 + \vec{q} + \vec{q}_{Jr}) + e^{i\varphi}u_{A_2}(\vec{K}_2 + \vec{q} + \vec{q}_{Jl})\right]$$

$$- \frac{J_\perp}{3J}\left[u_{B_2}(\vec{K}_2 + \vec{q} + \vec{q}_b) + e^{i\varphi}u_{B_2}(\vec{K}_2 + \vec{q} + \vec{q}_{Jr}) + e^{i\varphi}u_{B_2}(\vec{K}_2 + \vec{q} + \vec{q}_{Jl})\right]$$

$$(3b)$$

$$\Omega\, u_{A_2}(\vec{K}_2 + \vec{q}) = \left[\Omega_0 + 3\sqrt{3}\frac{D}{J}\right] u_{A_2}(\vec{K}_2 + \vec{q}) + \frac{\sqrt{3}a}{2}|\vec{q}|e^{-i(\theta_q + \theta/2)}u_{B_2}(\vec{K}_2 + \vec{q})$$

$$- \frac{J_\perp}{3J}\left[u_{A_1}(\vec{K}_1 + \vec{q} + \vec{q}_b) + e^{-i\varphi}u_{A_1}(\vec{K}_1 + \vec{q} + \vec{q}_{Jr}) + e^{i\varphi}u_{A_1}(\vec{K}_1 + \vec{q} + \vec{q}_{Jl})\right]$$

$$- \frac{J_\perp}{3J}\left[u_{B_1}(\vec{K}_1 + \vec{q} + \vec{q}_b) + e^{i\varphi}u_{B_1}(\vec{K}_1 + \vec{q} + \vec{q}_{Jr}) + e^{-i\varphi}u_{B_1}(\vec{K}_1 + \vec{q} + \vec{q}_{Jl})\right]$$

$$(3c)$$

$$\Omega\, u_{B_2}(\vec{K}_2 + \vec{q}) = \left[\Omega_0 - 3\sqrt{3}\,\frac{D}{J}\right] u_{B_2}(\vec{K}_2 + \vec{q}) + \frac{\sqrt{3}a}{2}|\vec{q}|\,e^{i(\theta_q + \theta/2)} u_{A_2}(\vec{K}_2 + \vec{q})$$

$$-\frac{J_\perp}{3J}\left[u_{A_1}(\vec{K}_1 + \vec{q} + \vec{q}_b) + u_{A_1}(\vec{K}_1 + \vec{q} + \vec{q}_{Jr}) + u_{A_1}(\vec{K}_1 + \vec{q} + \vec{q}_{Jl})\right]$$

$$-\frac{J_\perp}{3J}\left[u_{B_1}(\vec{K}_1 + \vec{q} + \vec{q}_b) + e^{-i\varphi} u_{B_1}(\vec{K}_1 + \vec{q} + \vec{q}_{Jr}) + e^{i\varphi} u_{B_1}(\vec{K}_1 + \vec{q} + \vec{q}_{Jl})\right]$$

$$(3d)$$

In equations 3, we have defined $\Omega = \frac{\omega}{JM}$, $\Omega_0 = 3 + \frac{2\tilde{J}_\perp(0)}{JA}$, $A = |\vec{a}_1 \times \vec{a}_2| = \sqrt{3}a^2/2$, and $\varphi = 2\pi/3$. The function $\tilde{J}_\perp(\vec{k})$ denotes the Fourier transform of the interlayer exchange coefficient. We have also defined the momenta

$$\vec{q}_b = \frac{8\pi \sin(\theta/2)}{3\sqrt{3}d}(0, -1)$$

$$\vec{q}_{Jr} = \frac{8\pi \sin(\theta/2)}{3\sqrt{3}d}(\sqrt{3}/2, 1/2)$$

$$\vec{q}_{Jl} = \frac{8\pi \sin(\theta/2)}{3\sqrt{3}d}(-\sqrt{3}/2, 1/2)$$

Equations 3 determine the system's Hamiltonian, $\mathcal{H}_T^K(\vec{q})$, near the $K-$ valley. The Hamiltonian $\mathcal{H}_T^{-K}(\vec{q})$ can then be deduced in a straight forward manner. Similar to the tBLG theory, the coupled amplitudes in the LL equations do not form a closed set. It is hence necessary to truncate the ensemble of LL equations involved in the formalism in order to construct $\mathcal{H}_T^K(\vec{q})$.

**Magnon magic angles and topological bands.** We fix the parameters $J_\perp = 0.2\,J$ and $\Omega_0 = 3.8$. The value of $\Omega_0$ only shifts the magnon spectrum and does not affect the main conclusions.

For negligible DMI, the Hamiltonian $\mathcal{H}_T^K(\vec{q})$ is qualitatively identical to the tBLG Hamiltonian and the magnons in tFBL mimic the electrons in tBLG. The first magic angle is found at $\theta \approx 1.8°$. The corresponding magnon band structure is presented in Fig.3a. The spectrum is calculated from both valley contributions and plotted along high symmetry axes in the moiré BZ (Fig.3g).

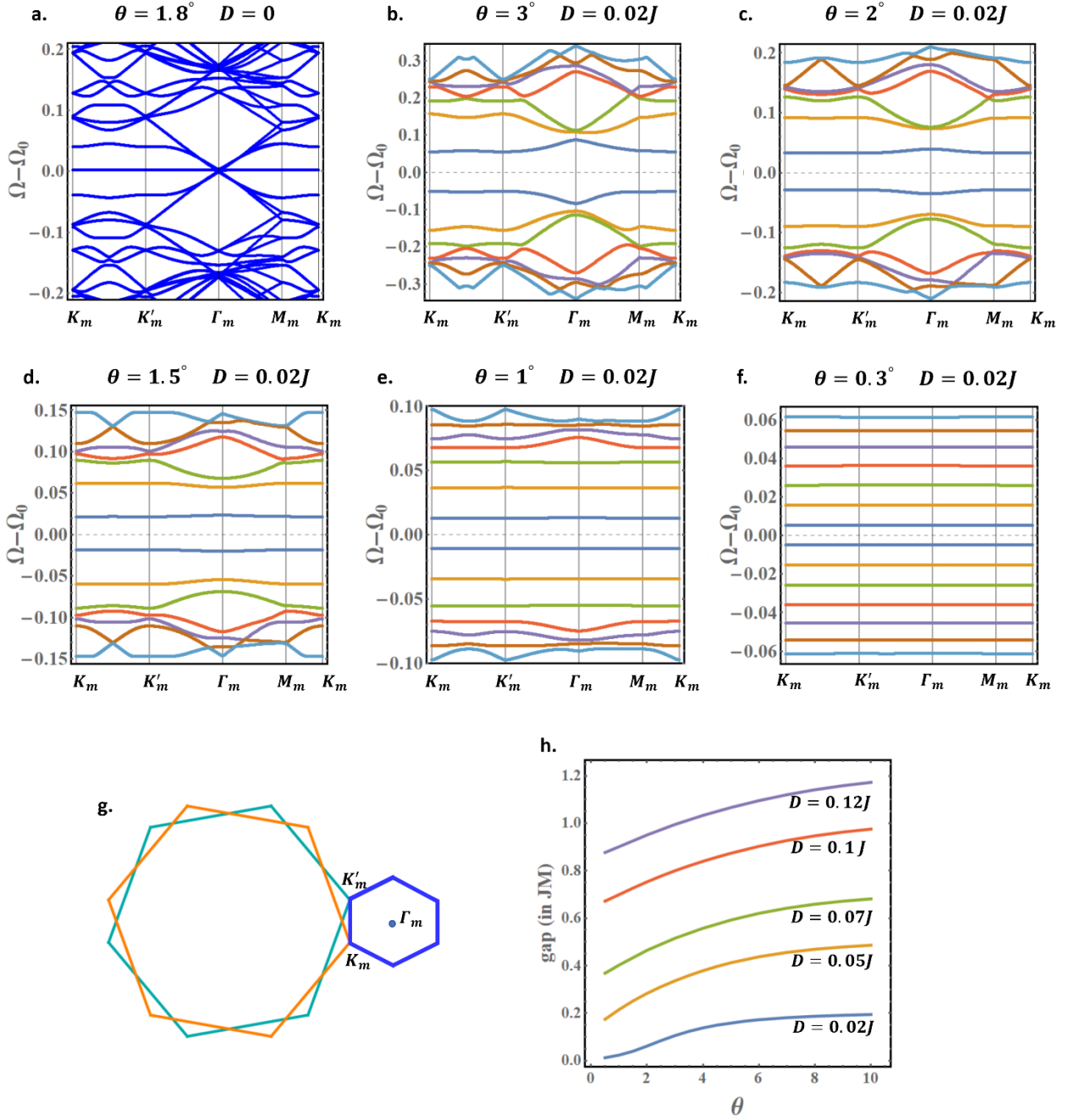

**Figure 3:** (a) Magic angle magnon spectrum for DMI-free tFBL. (b-f) Reconstruction of the $K-$ valley magnon spectrum for a tFBL with weak DMI ($D = 0.02\,J$). At slight twists, the spectrum presents multiple flat bands. (g) The rotated Brillouin zones for the 2 layers and the moiré BZ. (h) Dependence of the primary gap on the twist angle.

Introducing the slightest DMI strongly affects the dispersion profiles. Figs.3b-3f illustrate the reconstruction of the $K-$ valley magnon bands, caused by the twist angle $\theta$, in tFBL with weak DMI ($D = 0.02\,J$). For clarity, we only present 14 magnon bands near the $\Omega = \Omega_0$ axis. Very similar behavior is observed for the $-K$ valley magnon spectrum (not presented).

We use the notation $\epsilon_{\mu,i}(\vec{q})$ to denote the energies of the magnon bands. $\mu$ takes the values $\pm$ in reference to the $\pm K-$ valleys respectively. The bands above $\Omega_0$ are denoted by $i = 1, 2, \ldots$ in ascending energy order, while the bands below $\Omega_0$ are denoted by $i = -1, -2, \ldots$ in descending energy order.

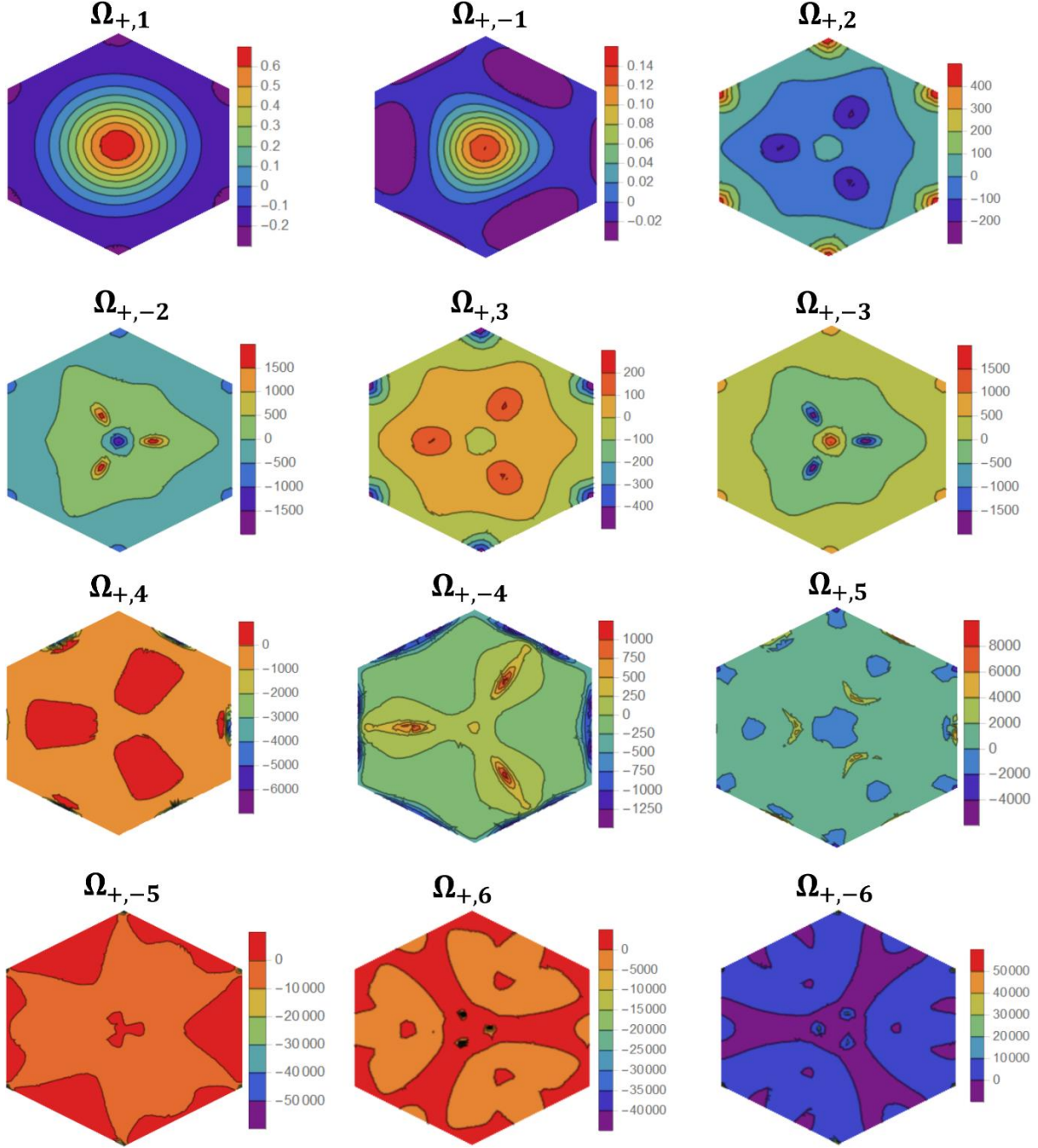

**Figure 4:** Berry curvatures plotted in the moiré BZ for selected $K-$ valley bands in a tFBL with $\theta = 3°$ and $D = 0.1J$.

The DMI induces a tunable primary energy gap between the valence-like band, $\epsilon_{+,-1}$, and the conduction-like band, $\epsilon_{+,1}$. The gap dependence on the twist angle $\theta$ and the DMI strength $D$ is analyzed in Fig. 3h.

The DMI also opens tiny gaps between neighboring bands $\epsilon_{\mu,i}$ and $\epsilon_{\mu,i+1}$. This decouples the bands and enables us to calculate their valley Berry curvatures, $\Omega_{\mu,i}$. As a sample, Figs.4 presents the $K-$ valley Berry curvatures for 12 bands ($\epsilon_{+,\pm i}, i = 1, ..., 6$) in a tFBL with $\theta = 3°$ and $D = 0.1J$. We have adopted the approach in [55] in our numerical calculation. The Berry curvatures, plotted in the moiré BZ, are peaked at avoided crossings between neighboring bands. Moreover, the Berry curvatures of topological bands display large values to compensate the reduced moiré BZ area.

**Table 1:** Chern numbers for selected bands, $\theta$, and $D$. In addition, $C_{+,\pm 1} = C_{+,\pm 2} = 0$ for these choices of $D$ and $\theta$.

| Chern number ($C_{\mu,i}$) | $D = 0.02\,J$ | | | | $D = 0.1\,J$ | | | |
|---|---|---|---|---|---|---|---|---|
| | $\theta = 4°$ | $\theta = 3.5°$ | $\theta = 3°$ | $\theta = 2.5°$ | $\theta = 4°$ | $\theta = 3.5°$ | $\theta = 3°$ | $\theta = 2.5°$ |
| $C_{+,3}$ | 3 | 0 | 0 | 0 | 0 | 0 | 0 | 0 |
| $C_{+,-3}$ | -3 | -3 | -3 | 0 | 0 | 0 | 0 | 0 |
| $C_{+,4}$ | 1 | 4 | 4 | 4 | -2 | -2 | -2 | -2 |
| $C_{+,-4}$ | -1 | -1 | -2 | -4 | -1 | -1 | -1 | -1 |
| $C_{+,5}$ | -1 | -1 | -1 | 1 | 4 | 4 | 4 | 4 |
| $C_{+,-5}$ | -1 | -1 | -1 | -1 | -3 | -3 | -2 | -1 |
| $C_{+,6}$ | 3 | 3 | 3 | 1 | 4 | 1 | -2 | -2 |
| $C_{+,-6}$ | -1 | -1 | -1 | -1 | -2 | 1 | 4 | 2 |

In the presence of the DMI, tFBL is topologically rich, presenting multiple magnon bands with nonzero Chern numbers $C_{\mu,i}$. Topological bands exist at any twist angle within the scope of the continuum approach ($\theta \leq 10°$). The corresponding Chern numbers are sensitive to the twist angle and the DMI strength. As an illustration, $K-$ valley Chern numbers are presented in Table 1 for selected bands, DMI and $\theta$. Generally, the $\pm K-$ valley Chern numbers can be deduced by the relation $C_{-K,i} = -C_{K,-i}$.

**Tunable magnon Hall and Nernst conductivities.** The non-trivial Berry curvatures and topological bands, consequences of the DMI, induce thermal magnon Hall and Nernst effects in tFBL. These effects exist in tFBL at any twist angle. Of particular interest, however, are the Hall and Nernst conductivities induces by the topological flat bands bundle. We choose tFBL with $D = 0.1J$ characterized by a bundle of (nearly) flat bands below $2^\circ$. Fig. 5a shows the first 12 flat bands $\epsilon_{+,\pm i}$, $i = 1, ..., 6$ for $\theta = 1.8^\circ$. The right panel illustrates the tiny gaps between these nearly flat bands. The nonzero Chern numbers for these 12 bands are investigated in Table 2 for the twist angle range $1.5^\circ \leq \theta \leq 2^\circ$. The table illustrates the strong and sensitive dependence of the Chern numbers on $\theta$. This naturally implies significant dependence of the Hall and Nernst conductivities on the twist angle.

**Table 2:** Illustrates the sensitivity of flat bands' Chern numbers to the twist angle in tFBL with $D = 0.1\,J$.

| Chern number $(C_{\mu,i})$ | $\theta = 2^\circ$ | $\theta = 1.9^\circ$ | $\theta = 1.8^\circ$ | $\theta = 1.7^\circ$ | $\theta = 1.6^\circ$ | $\theta = 1.5^\circ$ |
|---|---|---|---|---|---|---|
| $C_{+,-4}$ | 0 | 2 | 0 | 0 | 0 | 0 |
| $C_{+,5}$ | 2 | 2 | 2 | 0 | 0 | 0 |
| $C_{+,-5}$ | -2 | -4 | -2 | -2 | -2 | -2 |
| $C_{+,6}$ | -2 | -2 | -2 | 0 | 0 | 0 |
| $C_{+,-6}$ | 2 | 2 | 2 | 2 | 2 | 2 |

The Hall and Nernst conductivities, $\kappa_{xy}$ and $\alpha_{xy}^s$ respectively, are calculated using the standard equations [4, 8, 56-59],

$$\kappa_{xy} = -\frac{k_B^2 T}{\hbar V} \sum_{\vec{q},i,\mu} c_2\left(g\left(\epsilon_{\mu,i}(\vec{q})\right)\right) \Omega_{\mu,i}(\vec{q})$$

$$\alpha_{xy}^s = \frac{k_B}{V} \sum_{\vec{q},i,\mu} c_1\left(g\left(\epsilon_{\mu,i}(\vec{q})\right)\right) \Omega_{\mu,i}(\vec{q})$$

Here $g\left(\epsilon_{\mu,i}\right) = \left[e^{\epsilon_{\mu,i}/k_B T} - 1\right]^{-1}$ is the Bose-Einstein distribution function, while $c_1(x) = (1 + x)\ln(1 + x) - x \ln x$, and $c_2(x) = (1 + x)\left[\ln\left(\frac{1+x}{x}\right)\right]^2 - (\ln x)^2 - 2\text{Li}_2(-x)$. The symbol $\text{Li}_2$ stands for the dilogarithm function.

Figs. 5b and 5c present the tunable magnon Hall and Nernst conductivities, plotted as a function of temperature, in the twist angle interval $1.5^\circ \leq \theta \leq 2^\circ$. We have included contributions from 24 bands, namely $\epsilon_{\pm,\pm i}$, with $i = 1, ..., 6$. The curves are normalized relative to the maximum value in the plot. For the selected values of the DMI and twist angles, the conductivities show a standard profile: they vanish at $T = 0$ (no thermal excitations), change exponentially for larger temperatures, and approach a constant value at very large temperatures. The figures also illustrate the significant impact of the twist angle on the conductivities. Changing $\theta$ affects the energies and the Berry curvatures of the bands, and eventually modifies the Hall and Nernst conductivities. The impact on the energy is well determined: the energy bands are compressed closer to $\Omega_0$ for smaller $\theta$. The variation of the Berry curvatures, however, does not follow a simple or predefined trend. This can lead to an unsteady behavior in the Hall and Nernst conductivities, even when $\theta$ is varied smoothly. This is manifested in Figs.5c and 5d. The Nernst conductivity increases steadily from $\theta = 1.5°$ to $1.7°$, followed by a relatively abrupt jump at $\theta = 1.8°$. $\alpha_{xy}^s$ then decreases at $\theta = 1.9°$ prior to a significant enhancement at $\theta = 2°$. A similar behavior is observed for the Hall conductivity.

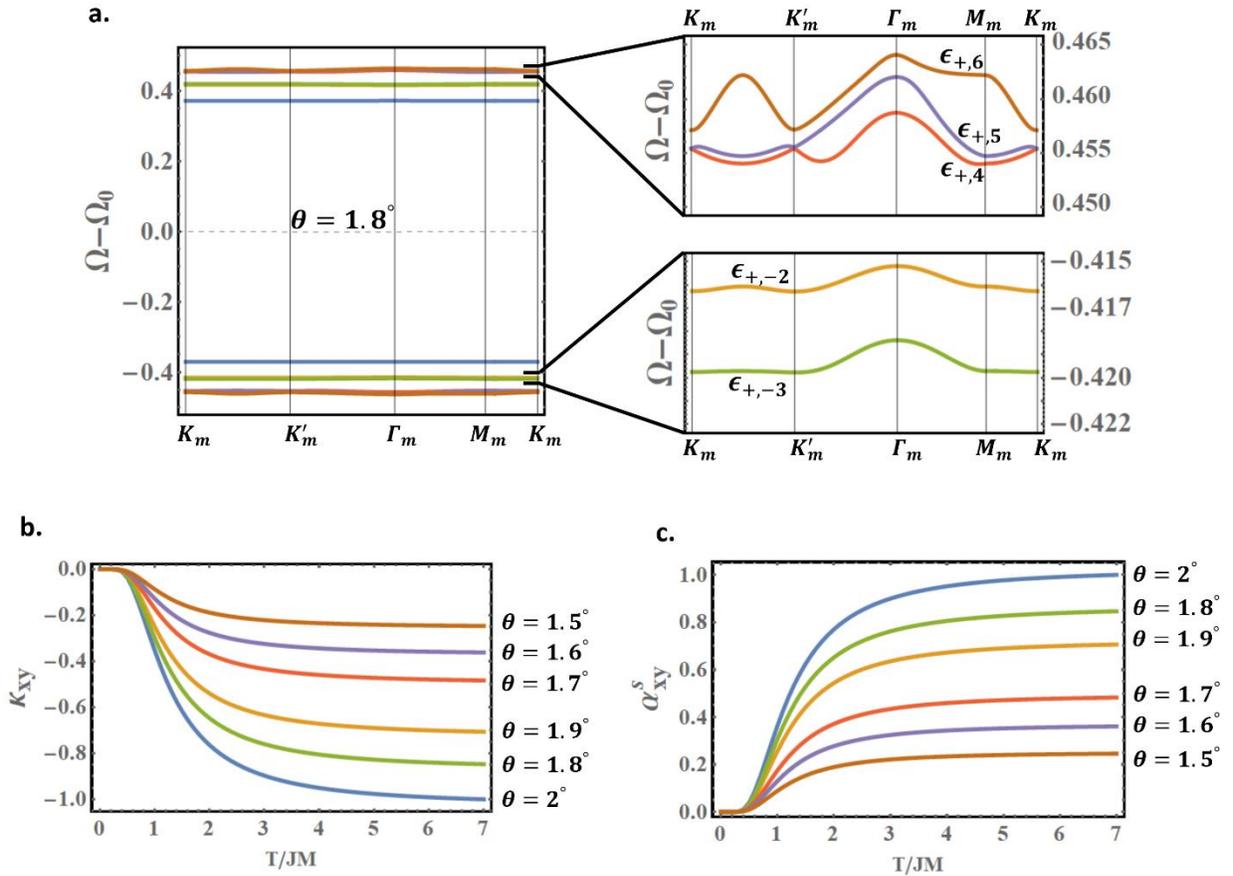

**Figure 5:** (a) Flat bands bundle in tFBL with $D = 0.1J$ and $\theta = 1.8°$. (b) and (c) show the tunable magnon Hall and Nernst conductivities, induced by the topological flat bands bundle, in tFBL with $D = 0.1J$.

## Conclusion

The present work proposes tFBL as a promising magnonic analogue for tBLG. In particular, we have focused on the topological physics induced by the DMI in tFBL.

The spin-orbit coupling is negligibly weak in tBLG. Nevertheless, magic angle flat bands in tBLG are topologically nontrivial [60-63], possibly due to the pseudo magnetic fields generated by the moiré potential [63]. Quantum anomalous Hall (QAH) effect was observed in magic angle tBLG on hexagonal Boron Nitride (hBN) substrate [64, 65, 66]. The Hall effect in tBLG, however, is present only at the magic angle and cannot be tuned through the twist angle.

Similar to tBLG, the twist angle in tFBL turns into a knob that can tune the magnon spectrum and consequently the magnetic properties of tFBL. Dzyaloshinskii-Moriya (DM) spin-orbit interaction, however, is present in 2D and quasi-2D magnets with broken inversion symmetry [6 - 8, 12, 14, 18, 21]. The DMI in tFBL induces topologically rich magnon bands for any twist angle. As a result, the DMI yields topological magnon Hall and Nernst conductivities that can be tuned via the twist angle. Edge states in tFBL are left for future studies. Unlike tBLG, tFBL with DMI presents a continuum of magic angles which might facilitate the experimental investigation of magnonic flat bands. In the lower part of the continuum, the magnon spectrum reconstructs in an unconventional manner, forming a bundle of topological flat bands. Experimental studies might reveal unconventional magic-angle phenomena similar to their fermionic counterparts.

Engineering magnon band gaps, flat bands, Nernst and Hall conductivities constitutes a difficult challenge for material science research. The ability to control all these characteristics via the twist angle in tFBL is indeed remarkable, and shall motivate interest in tFBL and its derivatives. Research on 2D moiré magnets is indeed active, with a current focus on twisted magnetic layers with antiferromagnetic interlayer coupling [67-70]. The collinear magnetic ground state in antiferromagnetically coupled layers is guaranteed only for specific ranges of the twist angle and material parameters [70]. These materials, however, are fundamentally different from the tFBL studied here, where the interlayer coupling is assumed ferromagnetic, like in $CrBr_3$ and $Cr_2Ge_2Te_6$. $Cr_2Ge_2Te_6$ is particularly interesting, as it is known to be a 2D Heisenberg ferromagnet with a slight out-of-plane magnetic anisotropy and a robust ferromagnetic interlayer coupling beyond the nearest neighbors [1, 71, 72]. With the rapidly growing family of 2D magnets, additional candidates for tFBL are likely to be discovered, opening novel horizons in the newly born field of 2D moiré magnets.

## Acknowledgments

Part of the numerical calculations was performed using the Phoenix High Performance Computing facility at the American University of the Middle East (AUM), Kuwait.

# Supplementary Information

# Magnon magic angles and tunable Hall conductivity in 2D twisted ferromagnetic bilayers

D. Ghader

**Supplementary Note 1. Spin waves in a honeycomb ferromagnetic monolayer.**

With the coordinate system presented in Fig.1a, the vectors $\vec{\delta}_i^A$ and $\vec{\gamma}_j$ are expressed as $\vec{\delta}_1^A = a(0, 1/\sqrt{3})$, $\vec{\delta}_2^A = a(1/2, -\sqrt{3}/6)$, $\vec{\delta}_3^A = a(-1/2, -\sqrt{3}/6)$, $\vec{\gamma}_1 = a(1/2, -\sqrt{3}/2)$, $\vec{\gamma}_2 = a(-1/2, -\sqrt{3}/2)$, $\vec{\gamma}_3 = a(1, 0)$, $\vec{\gamma}_4 = -\vec{\gamma}_1$, $\vec{\gamma}_5 = -\vec{\gamma}_2$, and $\vec{\gamma}_6 = -\vec{\gamma}_3$.

The lattice basis vectors are $\vec{a}_1 = a\left(\frac{1}{2}, \frac{\sqrt{3}}{2}\right)$ and $\vec{a}_2 = a\left(-1/2, \sqrt{3}/2\right)$ whereas the basis vectors in momentum space are $\vec{b}_1 = \frac{2\pi}{3d}(\sqrt{3}, 1)$ and $\vec{b}_2 = \frac{2\pi}{3d}(-\sqrt{3}, 1)$.

The semi-classical Heisenberg Hamiltonian including nearest neighbors exchange interaction and DMI can be expressed as

$$
\mathcal{H}_{ML} = -J \sum_{\vec{R}_A, \vec{\delta}_i^A} \vec{S}^A(\vec{R}_A, t) \cdot \vec{S}^B(\vec{R}_A + \vec{\delta}_i^A, t)
$$

$$
+ \sum_{\vec{R}_A, \vec{\gamma}_j} \vec{D}(\vec{R}_A, \vec{R}_A + \vec{\gamma}_j) \cdot \left[\vec{S}^A(\vec{R}_A, t) \times \vec{S}^A(\vec{R}_A + \vec{\gamma}_j, t)\right]
$$

$$
+ \sum_{\vec{R}_B, \vec{\gamma}_j} \vec{D}(\vec{R}_B, \vec{R}_B + \vec{\gamma}_j) \cdot \left[\vec{S}^B(\vec{R}_B, t) \times \vec{S}^B(\vec{R}_B + \vec{\gamma}_j, t)\right]
$$

The parameter $J$ denotes the in-plane exchange interaction coefficient. $\vec{D}$ is the alternating next nearest-neighbor DMI vector, with $\vec{D}(\vec{r}, \vec{r} + \vec{\gamma}_j) = \pm D\hat{z}$. The parameter $D$ denotes the strength of the DMI.

Defining the vectors $\vec{S}_D^{A/B} = S_y^{A/B}\hat{x} - S_x^{A/B}\hat{y}$ , we can re-write $\mathcal{H}_{ML}$ as

$$\mathcal{H}_{ML} = -J \sum_{\vec{R}_A, \vec{\delta}_i^A} \vec{S}^A(\vec{R}_A, t) . \vec{S}^B(\vec{R}_A + \vec{\delta}_i^A, t)$$

$$+ \sum_{\vec{R}_A, \vec{\gamma}_j} D_z(\vec{R}_A, \vec{R}_A + \vec{\gamma}_j) \vec{S}^A(\vec{R}_A, t) . \vec{S}_D^A(\vec{R}_A + \vec{\gamma}_j, t)$$

$$+ \sum_{\vec{R}_B, \vec{\gamma}_j} D_z(\vec{R}_B, \vec{R}_B + \vec{\gamma}_j) \vec{S}^B(\vec{R}_B, t) . \vec{S}_D^B(\vec{R}_B + \vec{\gamma}_j, t)$$

The effective fields $\vec{H}^A$ and $\vec{H}^B$ on A and B sites can then be written in terms of the magnetizations as

$$\vec{H}^{A/B}(\vec{R}_{A/B}, t) = -J \sum_{\vec{\delta}_i^{A/B}} \vec{M}^{B/A}(\vec{R}_{A/B} + \vec{\delta}_i^{A/B}, t) + \sum_{\vec{\gamma}_j} D_z(\vec{R}_{A/B}, \vec{R}_{A/B} + \vec{\gamma}_j) \vec{M}_D^{A/B}(\vec{R}_{A/B} + \vec{\gamma}_j, t)$$

Assuming plane wave solutions for $\vec{M}^{B/A}(\vec{r}, t)$ of the form

$$\vec{M}^{A/B}(\vec{r}, t) = \vec{M}_\parallel^{A/B}(\vec{r}, t) + M\hat{z} = \left( M_x^{A/B}\hat{x} + M_y^{A/B}\hat{y} \right) e^{i(wt + \vec{k}.\vec{r})} + M\hat{z}$$

we arrive at

$$\vec{H}^A(\vec{R}_A, t) = -3JM\hat{z} - J f(\vec{k}) \vec{M}_\parallel^B(\vec{R}_A, t) + D f_D(\vec{k}) \vec{M}_D^A(\vec{R}_A, t)$$

$$\vec{H}^B(\vec{R}_B, t) = -3JM\hat{z} - J f^*(\vec{k}) \vec{M}_\parallel^A(\vec{R}_B, t) + D f_D(\vec{k}) \vec{M}_D^B(\vec{R}_B, t)$$

with

$$f(\vec{k}) = e^{ik_y \frac{a}{\sqrt{3}}} + 2e^{-i\frac{\sqrt{3}a}{6}k_y} \cos\left(\frac{a}{2}k_x\right)$$

$$f_D(\vec{k}) = 4 i \sin\left(\frac{a}{2}k_x\right) \cos\left(\frac{\sqrt{3}a}{2}k_y\right) - 2 i \sin(k_x a)$$

Substituting in the LL equations of motion, $\partial_t \vec{M}^A(\vec{R}_A, t) = \vec{M}^A(\vec{R}_A, t) \times \vec{H}^A(\vec{R}_A, t)$, yields

$$i\omega M_x^A = -3JM M_y^A + JMf(\vec{k})M_y^B + DMf_D(\vec{k})M_x^A$$

$$i\omega M_y^A = 3JM M_x^A - JMf(\vec{k})M_x^B + DMf_D(\vec{k})M_y^A$$

Multiply the first equation by $-i$ and summing implies

$$\omega M^A = \left[3JM - iDMf_D(\vec{k})\right]M^A - JMf(\vec{k})M^B$$

In a similar way, the LL equation for the B-sublattice yields

$$\omega M^B = -JMf^*(\vec{k})M^A + \left[3JM + iDMf_D(\vec{k})\right]M^B$$

We hence arrive at the Hamiltonian for the ferromagnetic monolayer in the reciprocal space as

$$\mathcal{H}_{ML}(\vec{k}) = JM \begin{pmatrix} 3 - \frac{iD}{J}f_D(\vec{k}) & -f(\vec{k}) \\ -f^*(\vec{k}) & 3 + \frac{iD}{J}f_D(\vec{k}) \end{pmatrix}.$$

The reciprocal space Hamiltonian $\mathcal{H}_{ML}(\vec{k})$ admits 2 eigenvalues, namely

$$\Omega_\pm(\vec{k}) = \frac{\omega_\pm(\vec{k})}{JM} = 3 \pm \sqrt{\left|f(\vec{k})\right|^2 - \left(\frac{D}{J}f_D(\vec{k})\right)^2}$$

$\Omega_\pm(\vec{k})$ correspond to the conduction-like and valence-like bands for magnons in the 2D honeycomb ferromagnetic monolayer.

**Supplementary Note 2. The Heisenberg Hamiltonian for tFBL in a less compact form.**

$$\mathcal{H}_T = -J \sum_{\vec{R}_{A_1}, \vec{\delta}_i^A} \vec{S}^{A_1}(\vec{R}_{A_1}, t) \cdot \vec{S}^{B_1}(\vec{R}_{A_1} + \vec{\delta}_i^A, t) - J \sum_{\vec{R}_{A_2}, \vec{\delta}_i^A} \vec{S}^{A_2}(\vec{R}_{A_2}, t) \cdot \vec{S}^{B_2}(\vec{R}_{A_2} + \vec{\delta}_i^A, t)$$

$$+ \sum_{\vec{R}_{A_1}, \vec{\gamma}_j} \vec{D}(\vec{R}_{A_1}, \vec{R}_{A_1} + \vec{\gamma}_j) \cdot \left[ \vec{S}^{A_1}(\vec{R}_{A_1}, t) \times \vec{S}^{A_1}(\vec{R}_{A_1} + \vec{\gamma}_j, t) \right]$$

$$+ \sum_{\vec{R}_{B_1}, \vec{\gamma}_j} \vec{D}(\vec{R}_{B_1}, \vec{R}_{B_1} + \vec{\gamma}_j) \cdot \left[ \vec{S}^{B_1}(\vec{R}_{B_1}, t) \times \vec{S}^{B_1}(\vec{R}_{B_1} + \vec{\gamma}_j, t) \right]$$

$$+ \sum_{\vec{R}_{A_2}, \vec{\gamma}_j} \vec{D}(\vec{R}_{A_2}, \vec{R}_{A_2} + \vec{\gamma}_j) \cdot \left[ \vec{S}^{A_2}(\vec{R}_{A_2}, t) \times \vec{S}^{A_2}(\vec{R}_{A_2} + \vec{\gamma}_j, t) \right]$$

$$+ \sum_{\vec{R}_{B_2}, \vec{\gamma}_j} \vec{D}(\vec{R}_{B_2}, \vec{R}_{B_2} + \vec{\gamma}_j) \cdot \left[ \vec{S}^{B_2}(\vec{R}_{B_2}, t) \times \vec{S}^{B_2}(\vec{R}_{B_2} + \vec{\gamma}_j, t) \right]$$

$$- \sum_{\vec{R}_{A_1}, \vec{R}_{A_2}} J_\perp(\vec{R}_{A_1}, \vec{R}_{A_2}) \, \vec{S}^{A_1}(\vec{R}_{A_1}, t) \cdot \vec{S}^{A_2}(\vec{R}_{A_2}, t) - \sum_{\vec{R}_{A_1}, \vec{R}_{B_2}} J_\perp(\vec{R}_{A_1}, \vec{R}_{B_2}) \, \vec{S}^{A_1}(\vec{R}_{A_1}, t) \cdot \vec{S}^{B_2}(\vec{R}_{B_2}, t)$$

$$- \sum_{\vec{R}_{B_1}, \vec{R}_{A_2}} J_\perp(\vec{R}_{B_1}, \vec{R}_{A_2}) \, \vec{S}^{B_1}(\vec{R}_{B_1}, t) \cdot \vec{S}^{A_2}(\vec{R}_{A_2}, t) - \sum_{\vec{R}_{B_1}, \vec{R}_{B_2}} J_\perp(\vec{R}_{B_1}, \vec{R}_{B_2}) \, \vec{S}^{B_1}(\vec{R}_{B_1}, t) \cdot \vec{S}^{B_2}(\vec{R}_{B_2}, t)$$

**Supplementary Note 3. Landau-Lifshitz equations in tFBL.**

In the tFBL, the real and momentum space basis vectors are denoted $\vec{a}_{l,\alpha}$ and $\vec{b}_{l,\alpha}$ respectively ($\alpha = A$ or $B$ and $l = 1$ or $2$). These can be expressed as ($R_\theta$ is a 2D anticlockwise rotation by $\theta$)

$$\vec{a}_{2,\alpha} = R_{\theta/2}(\vec{a}_\alpha + \vec{\tau}_0)$$

$$\vec{a}_{1,\alpha} = R_{-\theta/2} \, \vec{a}_\alpha$$

$$\vec{b}_{2,\alpha} = R_{\theta/2} \, \vec{b}_\alpha$$

$$\vec{b}_{1,\alpha} = R_{-\theta/2} \, \vec{b}_\alpha$$

The positions of the atoms on the four sublattices can then be expressed as

$$\vec{R}_{A_1} = \vec{R}_1 + \vec{\tau}_{1,A}$$

$$\vec{R}_{B_1} = \vec{R}_1 + \vec{\tau}_{1,B}$$

$$\vec{R}_{A_2} = \vec{R}_2 + \vec{\tau}_{2,A}$$

$$\vec{R}_{B_2} = \vec{R}_2 + \vec{\tau}_{2,B}$$

with $\vec{R}_l = n_1 \vec{a}_{l,1} + n_2 \vec{a}_{l,2}$ $(n_1, n_2 \, \epsilon \, \mathbb{Z})$, $\vec{\tau}_{1,A} = (0,0)$, $\vec{\tau}_{1,B} = R_{-\theta/2}(0,d)$, $\vec{\tau}_{2,A} = R_{\theta/2}[(0,-d) + \vec{\tau}_0]$, and $\vec{\tau}_{2,B} = R_{\theta/2} \, \vec{\tau}_0$.

The twist gives rise to a moiré superlattice, with reciprocal basis vectors

$$\vec{b}_1^m = \vec{b}_{1,1} - \vec{b}_{2,1} = \frac{8\pi \sin(\theta/2)}{3d}(1, -\sqrt{3})$$

$$\vec{b}_2^m = \vec{b}_{1,2} - \vec{b}_{2,2} = \frac{8\pi \sin(\theta/2)}{3d}(1, \sqrt{3})$$

We recall the expression of the Heisenberg Hamiltonian $\mathcal{H}_T$ for the tFBL,

$$\mathcal{H}_T = -J \sum_{l, \vec{\delta}_i^A} \vec{S}^{A_l}(\vec{R}_{A_l}, t) . \vec{S}^{B_l}(\vec{R}_{A_l} + \vec{\delta}_i^A, t) - \sum_{\alpha, \beta} J_\perp(\vec{R}_{\alpha_1}, \vec{R}_{\beta_2}) \, \vec{S}^{\alpha_1}(\vec{R}_{\alpha_1}, t) . \vec{S}^{\beta_2}(\vec{R}_{\beta_2}, t)$$

$$+ \sum_{\alpha, l, \vec{\gamma}_j} \vec{D}(\vec{R}_{\alpha_l}, \vec{R}_{\alpha_l} + \vec{\gamma}_j) . [ \vec{S}^{\alpha_l}(\vec{R}_{\alpha_l}, t) \times \vec{S}^{\alpha_l}(\vec{R}_{\alpha_l} + \vec{\gamma}_j, t)]$$

(S1)

Similar to the monolayer case, the DMI term in $\mathcal{H}_T$ can be re-written as a scalar product,

$$\sum_{\vec{R}_{\alpha_l}, \vec{\gamma}_j} \vec{D}(\vec{R}_{\alpha_l}, \vec{R}_{\alpha_l} + \vec{\gamma}_j) . [ \vec{S}^{\alpha_l}(\vec{R}_{\alpha_l}, t) \times \vec{S}^{\alpha_l}(\vec{R}_{\alpha_l} + \vec{\gamma}_j, t)] =$$

$$\sum_{\vec{R}_{\alpha_l}, \vec{\gamma}_j} D_z(\vec{R}_{\alpha_l}, \vec{R}_{\alpha_l} + \vec{\gamma}_j) \vec{S}^{\alpha_l}(\vec{R}_{\alpha_l}, t) . \vec{S}_D^{\alpha_l}(\vec{R}_{\alpha_l} + \vec{\gamma}_j, t)$$

with $\vec{S}_D^{\alpha_l} = S_y^{\alpha_l} \hat{x} - S_x^{\alpha_l} \hat{y}$ .

We can now deduce the effective exchange fields $\vec{H}^{\alpha_l}$ acting on the magnetization $\vec{M}^{\alpha_l}$

$$\vec{H}^{\alpha_l}(\vec{R}_{\alpha_l},t) = -J\sum_{\vec{\delta}_i^\alpha}\vec{M}^{\bar{\alpha}_l}(\vec{R}_{\alpha_l}+\vec{\delta}_i^\alpha,t) + \sum_{\vec{\gamma}_j}D_z(\vec{R}_{\alpha_l},\vec{R}_{\alpha_l}+\vec{\gamma}_j)\vec{M}_D^{\alpha_l}(\vec{R}_{\alpha_l}+\vec{\gamma}_j,t)$$

$$-\sum_{\vec{R}_{\alpha_{\bar{l}}}}J_\perp(\vec{R}_{\alpha_l},\vec{R}_{\alpha_{\bar{l}}})\,\vec{M}^{\alpha_l}(\vec{R}_{\alpha_{\bar{l}}},t) - \sum_{\vec{R}_{\bar{\alpha}_{\bar{l}}}}J_\perp(\vec{R}_{\alpha_l},\vec{R}_{\bar{\alpha}_{\bar{l}}})\,\vec{M}^{\bar{\alpha}_l}(\vec{R}_{\bar{\alpha}_{\bar{l}}},t)$$

(S2)

where we have used the convention that if $\alpha = A$ then $\bar{\alpha} = B$ and vice versa. Same convention assumed for $l$ and $\bar{l}$.

We assume harmonic time dependence (with frequency $\omega$) for the magnetizations. The $x$ and $y$ components of the Landau-Lifshitz (LL) equations of motion, $\partial_t\vec{M}^{\alpha_l} = \vec{M}^{\alpha_l}\times\vec{H}^{\alpha_l}$, yield 2 equations of motion for each sublattice $\alpha_l$. Combining the $x$ and $y$ equations yield

$$\omega M^{\alpha_l}(\vec{R}_{\alpha_l}) = \left[3JM + M\sum_{\vec{R}_{\alpha_{\bar{l}}}}J_\perp(\vec{R}_{\alpha_l},\vec{R}_{\alpha_{\bar{l}}}) + M\sum_{\vec{R}_{\bar{\alpha}_{\bar{l}}}}J_\perp(\vec{R}_{\alpha_l},\vec{R}_{\bar{\alpha}_{\bar{l}}})\right]M^{\alpha_l}(\vec{R}_{\alpha_l})$$

$$-JM\sum_{\vec{\delta}_i^\alpha}M^{\bar{\alpha}_l}(\vec{R}_{\alpha_l}+\vec{\delta}_i^\alpha) - iM\sum_{\vec{\gamma}_j}D_z(\vec{R}_{\alpha_l},\vec{R}_{\alpha_l}+\vec{\gamma}_j)M^{\alpha_l}(\vec{R}_{\alpha_l}+\vec{\gamma}_j,t)$$

$$-M\sum_{\vec{R}_{\alpha_{\bar{l}}}}J_\perp(\vec{R}_{\alpha_l},\vec{R}_{\alpha_{\bar{l}}})\,M^{\alpha_l}(\vec{R}_{\alpha_{\bar{l}}}) - M\sum_{\vec{R}_{\bar{\alpha}_{\bar{l}}}}J_\perp(\vec{R}_{\alpha_l},\vec{R}_{\bar{\alpha}_{\bar{l}}})\,M^{\bar{\alpha}_l}(\vec{R}_{\bar{\alpha}_{\bar{l}}})$$

(S3)

with $M^{\alpha_l} = M_x^{\alpha_l} + iM_y^{\alpha_l}$.

We next expand the magnetization amplitudes in terms of Bloch waves

$$\frac{\omega}{\sqrt{N_l}}\sum_{\vec{k}_l'}e^{i\vec{k}_l'\cdot\vec{R}_{\alpha_l}}u_{\alpha_l}(\vec{k}_l') = -\frac{JM}{\sqrt{N_l}}\sum_{\vec{k}_l'}f_{ex}^\alpha(\vec{k}_l')\,e^{i\vec{k}_l'\cdot\vec{R}_{\alpha_l}}\,u_{\bar{\alpha}_l}(\vec{k}_l') +$$

$$\frac{1}{\sqrt{N_l}}\left[3MJ + MDf_D^\alpha(\vec{k}_l') + M\sum_{\vec{R}_{\alpha_{\bar{l}}}}J_\perp(\vec{R}_{\alpha_l},\vec{R}_{\alpha_{\bar{l}}}) + M\sum_{\vec{R}_{\bar{\alpha}_{\bar{l}}}}J_\perp(\vec{R}_{\alpha_l},\vec{R}_{\bar{\alpha}_{\bar{l}}})\right]\sum_{\vec{k}_l'}e^{i\vec{k}_l'\cdot\vec{R}_{\alpha_l}}u_{\alpha_l}(\vec{k}_l')$$

$$-\frac{M}{\sqrt{N_{\bar{l}}}}\sum_{\vec{R}_{\alpha_{\bar{l}}},\vec{k}_{\bar{l}}}J_\perp(\vec{R}_{\alpha_l},\vec{R}_{\alpha_{\bar{l}}})e^{i\vec{k}_{\bar{l}}\cdot\vec{R}_{\alpha_{\bar{l}}}}u_{\alpha_l}(\vec{k}_{\bar{l}}) - \frac{M}{\sqrt{N_{\bar{l}}}}\sum_{\vec{R}_{\bar{\alpha}_{\bar{l}}},\vec{k}_{\bar{l}}}J_\perp(\vec{R}_{\alpha_l},\vec{R}_{\bar{\alpha}_{\bar{l}}})e^{i\vec{k}_{\bar{l}}\cdot\vec{R}_{\bar{\alpha}_{\bar{l}}}}u_{\bar{\alpha}_{\bar{l}}}(\vec{k}_{\bar{l}})$$

(S4)

$N_l$ and $N_{\bar{l}}$ are the number of unit cells while $\vec{k}'_l$ and $\vec{k}_{\bar{l}}$ are wave vectors in layers $l$ and $\bar{l}$. We have also defined

$$f^A_{ex}(\vec{k}'_l) = \sum_{\vec{\delta}^A_l} e^{i\vec{k}'_l \cdot \vec{\delta}^A_l} = e^{ik'_{l,y}\frac{a}{\sqrt{3}}} + 2e^{-i\frac{\sqrt{3}a}{6}k'_{l,y}} \cos\left(\frac{a}{2}k'_{l,x}\right) = \left(f^B_{ex}(\vec{k}'_l)\right)^*$$

$$f^A_D(\vec{k}'_l) = \sum_{\vec{\gamma}_j} e^{i\vec{k}'_l \cdot \vec{\gamma}_j} = 4\sin\left(\frac{a}{2}k_x\right)\cos\left(\frac{\sqrt{3}a}{2}k_y\right) - 2\sin(k_x a) = -f^B_D(\vec{k}'_l)$$

$$\text{(S5)}$$

Finally, we multiply equation S4 by $e^{-i\vec{k}_l \cdot \vec{R}_{\alpha_l}}$ and sum the whole equation over $\vec{R}_{\alpha_l}$ to get

$$\omega\, u_{\alpha_l}(\vec{k}_l) = [3MJ + MDf^\alpha_D(\vec{k}'_l)]u_{\alpha_l}(\vec{k}_l) - JMf^\alpha_{ex}(\vec{k}'_l)u_{\bar{\alpha}_l}(\vec{k}_l)$$

$$+ M\sum_{\vec{k}'_l}[\mathcal{J}^{\alpha_l,\alpha_l}(\vec{k}_l,\vec{k}'_l) + \mathcal{J}^{\alpha_l,\bar{\alpha}_l}(\vec{k}_l,\vec{k}'_l)]\,u_{\alpha_l}(\vec{k}'_l)$$

$$- M\sum_{\vec{k}_{\bar{l}}}\mathcal{J}^{\alpha_l,\alpha_l}_\perp(\vec{k}_l,\vec{k}_{\bar{l}})\,u_{\alpha_l}(\vec{k}_{\bar{l}}) - M\sum_{\vec{k}_{\bar{l}}}\mathcal{J}^{\alpha_l,\bar{\alpha}_l}_\perp(\vec{k}_l,\vec{k}_{\bar{l}})\,u_{\bar{\alpha}_l}(\vec{k}_{\bar{l}})$$

$$\text{(S6)}$$

with the interlayer coefficients defined as

$$\mathcal{J}^{\alpha_l,\alpha_{\bar{l}}}_\perp(\vec{k}_l,\vec{k}_{\bar{l}}) = \frac{1}{\sqrt{N_l N_{\bar{l}}}}\sum_{\vec{R}_{\alpha_l},\vec{R}_{\alpha_{\bar{l}}}} e^{-i\vec{k}_l \cdot \vec{R}_{\alpha_l}} J_\perp(\vec{R}_{\alpha_l},\vec{R}_{\alpha_{\bar{l}}}) e^{i\vec{k}_{\bar{l}} \cdot \vec{R}_{\alpha_{\bar{l}}}}$$

$$\text{(S7a)}$$

$$\mathcal{J}^{\alpha_l,\bar{\alpha}_{\bar{l}}}_\perp(\vec{k}_l,\vec{k}_{\bar{l}}) = \frac{1}{\sqrt{N_l N_{\bar{l}}}}\sum_{\vec{R}_{\alpha_l},\vec{R}_{\bar{\alpha}_{\bar{l}}}} e^{-i\vec{k}_l \cdot \vec{R}_{\alpha_l}} J_\perp(\vec{R}_{\alpha_l},\vec{R}_{\bar{\alpha}_{\bar{l}}}) e^{i\vec{k}_{\bar{l}} \cdot \vec{R}_{\bar{\alpha}_{\bar{l}}}}$$

$$\text{(S7b)}$$

while the intralayer coefficients read

$$\mathcal{J}^{\alpha_l,\alpha_l}(\vec{k}_l,\vec{k}'_l) = \frac{1}{N_l}\sum_{\vec{R}_{\alpha_l},\vec{R}_{\alpha_{\bar{l}}}} e^{-i(\vec{k}_l - \vec{k}'_l)\cdot\vec{R}_{\alpha_l}} J_\perp(\vec{R}_{\alpha_l},\vec{R}_{\alpha_{\bar{l}}})$$

$$\text{(S8a)}$$

$$\mathcal{J}^{\alpha_l \bar{\alpha}_l}(\vec{k}_l, \vec{k}_l') = \frac{1}{N_l} \sum_{\vec{R}_{\alpha_l}, \vec{R}_{\bar{\alpha}_l}} e^{-i(\vec{k}_l - \vec{k}_l') \cdot \vec{R}_{\alpha_l}} J_{\perp}(\vec{R}_{\alpha_l}, \vec{R}_{\bar{\alpha}_l})$$

(S8b)

The interlayer terms in the LL equations are qualitatively identical to those encountered in the electronic theory of tBLG. The Bistritzer - MacDonald continuum approach yields the identities

$$\mathcal{J}_{\perp}^{\alpha_l, \alpha_{\bar{l}}}(\vec{K}_l + \vec{q}_l, \vec{K}_{\bar{l}} + \vec{q}_{\bar{l}}) = \frac{J_{\perp}}{3} \Big[ \delta_{\vec{q}_l - \vec{q}_{\bar{l}}, -(\vec{K}_l - \vec{K}_{\bar{l}})} + e^{i\vec{b}_{\bar{l},2} \cdot \vec{\tau}_{l,\alpha}} \ e^{-i\vec{b}_{\bar{l},2} \cdot \vec{\tau}_{\bar{l},\alpha}} \ \delta_{\vec{q}_l - \vec{q}_{\bar{l}}, -(\vec{K}_l - \vec{K}_{\bar{l}} + \vec{b}_{l,2} - \vec{b}_{\bar{l},2})}$$
$$+ e^{-i\vec{b}_{l,1} \cdot \vec{\tau}_{l,\alpha}} \ e^{i\vec{b}_{\bar{l},1} \cdot \vec{\tau}_{\bar{l},\alpha}} \ \delta_{\vec{q}_l - \vec{q}_{\bar{l}}, -(\vec{K}_l - \vec{K}_{\bar{l}} - \vec{b}_{l,1} + \vec{b}_{\bar{l},1})} \Big]$$

(S9a)

$$\mathcal{J}_{\perp}^{\alpha_l, \bar{\alpha}_{\bar{l}}}(\vec{K}_l + \vec{q}_l, \vec{K}_{\bar{l}} + \vec{q}_{\bar{l}}) = \frac{J_{\perp}}{3} \Big[ \delta_{\vec{q}_l - \vec{q}_{\bar{l}}, -(\vec{K}_l - \vec{K}_{\bar{l}})} + e^{i\vec{b}_{\bar{l},2} \cdot \vec{\tau}_{l,\alpha}} \ e^{-i\vec{b}_{\bar{l},2} \cdot \vec{\tau}_{\bar{l},\bar{\alpha}}} \ \delta_{\vec{q}_l - \vec{q}_{\bar{l}}, -(\vec{K}_l - \vec{K}_{\bar{l}} + \vec{b}_{l,2} - \vec{b}_{\bar{l},2})}$$
$$+ e^{-i\vec{b}_{l,1} \cdot \vec{\tau}_{l,\alpha}} \ e^{i\vec{b}_{\bar{l},1} \cdot \vec{\tau}_{\bar{l},\bar{\alpha}}} \ \delta_{\vec{q}_l - \vec{q}_{\bar{l}}, -(\vec{K}_l - \vec{K}_{\bar{l}} - \vec{b}_{l,1} + \vec{b}_{\bar{l},1})} \Big]$$

(S9b)

We now consider the intralayer coefficients presented in equations S8c and S8d, absent in the electronic theory of graphene. The starting point is the Fourier transform of $J_{\perp}(\vec{R}_{\alpha_l}, \vec{R}_{\alpha_{\bar{l}}})$,

$$\mathcal{J}^{\alpha_l, \alpha_{\bar{l}}}(\vec{k}_l, \vec{k}_l') = \frac{1}{N_l} \sum_{\vec{R}_{\alpha_l}, \vec{R}_{\alpha_{\bar{l}}}} e^{-i(\vec{k}_l - \vec{k}_l') \cdot \vec{R}_{\alpha_l}} J_{\perp}(\vec{R}_{\alpha_l}, \vec{R}_{\alpha_{\bar{l}}})$$

$$= \frac{1}{N_l} \int_{\mathbb{R}^2} \frac{d^2 \vec{p}}{(2\pi)^2} \tilde{J}_{\perp}(\vec{p}) \sum_{\vec{R}_l} e^{-i(\vec{k}_l - \vec{k}_l' - \vec{p}) \cdot (\vec{R}_l + \vec{\tau}_{l,\alpha})} \sum_{\vec{R}_{\bar{l}}} e^{-i\vec{p} \cdot (\vec{R}_l + \vec{\tau}_{l,\alpha})}$$

$$= N_{\bar{l}} \int_{\mathbb{R}^2} \frac{d^2 \vec{p}}{(2\pi)^2} \tilde{J}_{\perp}(\vec{p}) \sum_{\vec{G}_l, \vec{G}_{\bar{l}}} e^{-i\vec{G}_l \cdot \vec{\tau}_{l,\alpha}} \ e^{-i\vec{G}_{\bar{l}} \cdot \vec{\tau}_{l,\alpha}} \ \delta_{\vec{k}_l - \vec{k}_l' - \vec{p}, \vec{G}_l} \ \delta_{\vec{p}, \vec{G}_{\bar{l}}}$$

$$= \frac{1}{A} \sum_{\vec{G}_l, \vec{G}_{\bar{l}}} \tilde{J}_{\perp}(\vec{G}_{\bar{l}}) e^{-i\vec{G}_l \cdot \vec{\tau}_{l,\alpha}} \ e^{i\vec{G}_{\bar{l}} \cdot \vec{\tau}_{l,\alpha}} \ \delta_{\vec{k}_l - \vec{k}_l', \vec{G}_l - \vec{G}_{\bar{l}}}$$

In the present case, both $\vec{k}_l$ and $\vec{k}_l'$ are expanded near $K_l$,

$$\mathcal{J}^{\alpha_l, \alpha_{\bar{l}}}(\vec{K}_l + \vec{q}_l, \vec{K}_l + \vec{q}_l') = \frac{1}{A} \sum_{\vec{G}_l, \vec{G}_{\bar{l}}} \tilde{J}_{\perp}(\vec{G}_{\bar{l}}) e^{-i\vec{G}_l \cdot \vec{\tau}_{l,\alpha}} \ e^{i\vec{G}_{\bar{l}} \cdot \vec{\tau}_{l,\alpha}} \ \delta_{\vec{q}_l - \vec{q}_l', \vec{G}_l - \vec{G}_{\bar{l}}}$$

(S10a)

Similarly

$$\mathcal{J}^{\alpha_l,\bar{\alpha}_{\bar{l}}}(\vec{K}_l+\vec{q}_l,\vec{K}_l+\vec{q}_l')=\frac{1}{A}\sum_{\vec{G}_l,\vec{G}_{\bar{l}}}\tilde{J}_\perp(\vec{G}_{\bar{l}})e^{-i\vec{G}_l\cdot\vec{\tau}_{l,\alpha}}\,e^{i\vec{G}_{\bar{l}}\cdot\vec{\tau}_{\bar{l},\bar{\alpha}}}\,\delta_{\vec{q}_l-\vec{q}_l',\vec{G}_l-\vec{G}_{\bar{l}}}$$

$$\text{(S10b)}$$

Near $K_l$, the vectors $\vec{q}_l-\vec{q}_l'$ in equations S10 are very small and match only moiré reciprocal lattice vectors $\vec{G}^m=\vec{G}_l-\vec{G}_{\bar{l}}=\pm(R_{-\theta/2}\,\vec{G}-R_{\theta/2}\,\vec{G})$. Here $\vec{G}=n_1\vec{b}_1+n_2\vec{b}_2$ is a reciprocal lattice vector of the unrotated honeycomb monolayer. The summation in S10 hence reduces to a summation over $\vec{G}$ of the unrotated lattice. For example,

$$\mathcal{J}^{A_1,A_2}(\vec{K}_1+\vec{q}_1,\vec{K}_1+\vec{q}_1')=\frac{1}{A}\sum_{\vec{G}}\tilde{J}_\perp(|\vec{G}|)e^{-i\vec{G}\cdot(0,0)}\,e^{i\vec{G}\cdot[(0,-d)+\vec{\tau}_0]}\,\delta_{\vec{q}_1-\vec{q}_1',R_{-\theta/2}\,\vec{G}-R_{\theta/2}\,\vec{G}}$$

$$\text{(S11)}$$

In the summation present in equation 11, we only need to consider the most relevant contributions, namely $\vec{G}=\vec{0},\pm\vec{b}_1$, and $\pm\vec{b}_2$. Consequently,

$$\mathcal{J}^{A_1,A_2}(\vec{K}_1+\vec{q}_1,\vec{K}_1+\vec{q}_1')=$$

$$\frac{\tilde{J}_\perp(0)}{A}\delta_{\vec{q}_1-\vec{q}_1',\vec{0}}+\frac{\tilde{J}_\perp(\sqrt{3}\times|\vec{K}|)}{A}\left[e^{i(\vec{b}_1\cdot\vec{\tau}_0-\varphi)}\delta_{\vec{q}_1-\vec{q}_1',\vec{G}_1^m}+e^{-i(\vec{b}_1\cdot\vec{\tau}_0-\varphi)}\delta_{\vec{q}_1-\vec{q}_1',-\vec{G}_1^m}\right]$$

$$+\frac{\tilde{J}_\perp(\sqrt{3}\times|\vec{K}|)}{A}\left[e^{i(\vec{b}_2\cdot\vec{\tau}_0-\varphi)}\delta_{\vec{q}_1-\vec{q}_1',\vec{G}_2^m}+e^{-i(\vec{b}_2\cdot\vec{\tau}_0-\varphi)}\delta_{\vec{q}_1-\vec{q}_1',-\vec{G}_2^m}\right]$$

with $\varphi=2\pi/3$, $\vec{G}_1^m=R_{-\theta/2}\,\vec{b}_1-R_{\theta/2}\,\vec{b}_1$ and $\vec{G}_2^m=R_{-\theta/2}\,\vec{b}_2-R_{\theta/2}\,\vec{b}_2$. We have also used the fact $\tilde{J}_\perp(|\vec{b}_1|)=\tilde{J}_\perp(|\vec{b}_2|)=\tilde{J}_\perp(\sqrt{3}\times|\vec{K}_1|)=\tilde{J}_\perp(\sqrt{3}\times|\vec{K}|)$.

Before proceeding, we note that for the case $\theta=0$, the summation in S11 becomes infinite and $\mathcal{J}^{\alpha_l,\alpha_{\bar{l}}}$ converges to $J_\perp(d_{\alpha_l,\alpha_{\bar{l}}})$, where $d_{\alpha_l,\alpha_{\bar{l}}}$ denotes the distance between sites $\alpha_l$ and $\alpha_{\bar{l}}$. This perfectly reproduces the AA/AB stacking cases.

In van der Waals magnetic materials, the interlayer Fourier transform $\tilde{J}_\perp(k)$ is extremely sharp and $\tilde{J}_\perp(\sqrt{3}\times|\vec{K}|)$ is negligible compared to $\tilde{J}_\perp(0)$. We hence arrive at the simple expressions

$$\mathcal{J}^{\alpha_l \alpha_{\bar{l}}}\left(\vec{K}_l + \vec{q}_l, \vec{K}_l + \vec{q}_l'\right) \approx \mathcal{J}^{\alpha_l \bar{\alpha}_{\bar{l}}}\left(\vec{K}_l + \vec{q}_l, \vec{K}_l + \vec{q}_l'\right) \approx \frac{J_\perp(0)}{A} \delta_{\vec{q}_1, \vec{q}_1'}$$

<div align="right">(S12)</div>

With this faithful approximation, the magnon theory is independent of $\vec{\tau}_0$ as in tBLG (we set $\vec{\tau}_0 = \vec{0}$). Substituting equations S9 and S12 in S6 then expanding $f_{ex}^\alpha(\vec{k}_l')$ and $f_D^\alpha$ near $K_l$ and $K_{\bar{l}}$ yields the final expressions of the LL equations (equations 3 in the main text).